\documentclass[aps,prl,reprint,superscriptaddress]{revtex4-2}
\usepackage{graphicx}  % needed for figures
\usepackage{dcolumn}   % needed for some tables
\usepackage{bm}        % for math
\usepackage{amssymb}   % for math
\usepackage{amsmath}
\usepackage{xcolor}
\begin{document}

\title{Optimizing Evolution Under Cyclic Selection: the Case of B Cells}

\author{Xingjian Hou}
\affiliation{Lewis-Sigler Institute for Integrative Genomics, Princeton University, Princeton, New Jersey 08544, USA}

\author{Colin Scheibner}
\email{cs5096@princeton.edu}
\affiliation{Joseph Henry Laboratories of Physics, Princeton University, Princeton, New Jersey 08544, USA}
\affiliation{Princeton Center for Theoretical Science, Princeton University, Princeton, New Jersey 08544, USA}

\author{Ned S. Wingreen}
\email{wingreen@princeton.edu}
\affiliation{Lewis-Sigler Institute for Integrative Genomics, Princeton University, Princeton, New Jersey 08544, USA}
\affiliation{Department of Molecular Biology, Princeton University, Princeton, New Jersey 08544, USA}

\begin{abstract}
Rapid evolutionary adaptation may be necessary for organisms facing strong environmental pressures. Although selection often acts continuously, in the case of B cells evolving high affinity antibodies in germinal centers, selection can take place after multiple rounds of division. To study how the number of divisions between selection events affects adaptation, we develop a minimal computational model of cyclic selection. We find that the optimal average mutation rate per division and the corresponding adaptation velocity are nearly invariant across number of divisions between selections, and we quantitatively explain this behavior with a theory based on the dynamics of the ``nose", i.e. the fittest individuals in the population. We also find that when the number of divisions between selection events is large, then mutating every division results in slower adaptation than mutating only at the last division before selection (as is the case for B cells). Our model provides a framework for understanding adaptation in natural and experimental systems subject to cyclic selection.\end{abstract}

\maketitle

Under strong environmental pressures, living systems may need to evolve rapidly to adapt and survive. In the adaptive immune system, for example, B cells quickly evolve their receptors to produce high-affinity antibodies that can neutralize antigens \cite{victora2012,DespondsPNAS2016,DeWittCell2026,NourmohammadMBE2019}. While many organisms are subject to constant selection, this is not necessarily the case for cells inside multicellular organisms, which can evolve under controlled cellular microenvironments. B cells in particular can divide multiple times before mutated progeny are selected \cite{Gitlin2014nature}. How does the interval between selection events affect adaptation?

Studies on population responses to intermittently selective environments have largely focused on the accumulation of deleterious mutations and the evolution of gene regulatory strategies \citep{WardlawGenetics2012,GerlandPNAS2009,GyorgyNatureComm2023}, while studies of positive adaptation are comparatively limited. One such study assumed an effectively monomorphic population, applicable when environmental changes occur on timescales longer than the fixation time of new mutations \citep{TrubenovaEvolution2019}. To derive more general theories applicable to B cells and other systems, we adopt an alternative approach based on fitness distributions to describe the evolution of a polymorphic population \cite{desai2007genetics,GoodPNAS2012,Rouzine2008TheoPopBio,PyoPRL2025}.

\begin{figure}
    \centering
    \includegraphics[width=\linewidth]{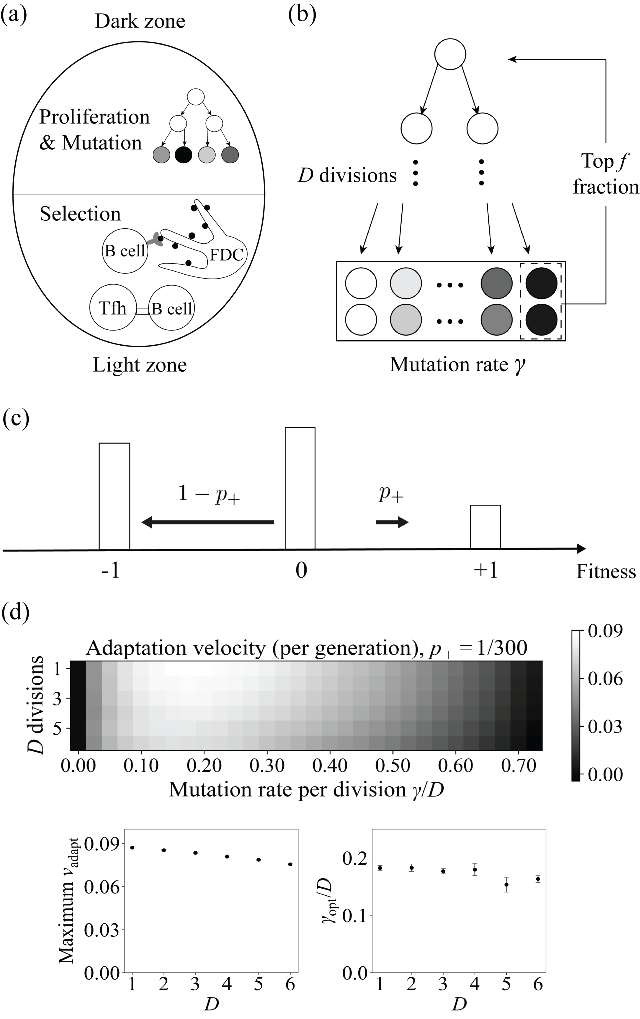}
    \caption{B cell affinity maturation processes and adaptation velocities. (a) B cells undergo proliferation and somatic hypermutation (SHM) in the dark zone of the germinal center, and undergo selection processes via interaction with FDCs and Tfh cells in the light zone. (b) Simplified model of B cell division, mutation, and selection. Circle colors represent phenotypes, with darker colors representing more fit individuals. (c) Schematic of positive and negative mutations having discrete $\pm 1$ effects on fitness, with positive mutations being rarer than negative mutations. (d) Heat map: Modeled adaptation rates for different numbers of divisions $D$ per selection cycle and mutation rates per division $\gamma /D$, for fixed maximal population $N=1024$ (the right order of magnitude for the B cell population in a germinal center \cite{Wittenbrinkjimmun2010,PelissierCells2020}). Bottom plots: Maximum adaptation velocities $v_\mathrm{adapt}$ and the corresponding optimal mutation rates per division $\gamma _\mathrm{opt}/D$ for different numbers of division $D$.}
    \label{fig:1}
\end{figure}

This evolutionary process of B cell affinity maturation takes place in germinal centers (Fig.~\ref{fig:1}(a)). In the dark zone of the germinal center, B cells undergo proliferation and somatic hypermutation (SHM) focused on the B cell receptor. The progeny B cells then migrate to the light zone, where they compete first for antigen presented on follicular dendritic cells (FDCs) and then for stimulating interactions with T follicular helper (Tfh) cells \cite{Allen2007Science,JiangPNAS2024}. B cells that present more cognate antigen to Tfh cells are stimulated to divide more times in the dark zone before their progeny return to the light zone for another round of selection \cite{victora2012,Gitlin2014nature}. 

Motivated by B cell affinity maturation in the germinal center, we study whether and how a system that only intermittently selects can still achieve a high rate of adaptive evolution. Perhaps surprisingly, we find that under conditions relevant to germinal centers, the optimal adaptation rate is nearly independent of selection frequency, and these optima are achieved with the same average rate of mutations \textit{per division} (not per selection). We account for this behavior with a theory focused on the dynamics of the ``nose'', i.e., the population’s highest-fitness individuals. For low selection frequencies, we find that mutating every division results in slower adaptation than mutating only at the last division before selection, suggesting that this observed property of B cells might accelerate their adaptation. 

To isolate the effects of selection frequency and mutation rate on adaptation velocity, we consider a minimal model of periodic selection (Fig.~\ref{fig:1}(b)).  
Motivated by the separation of light and dark zones in germinal centers, we consider agents that undergo proliferation and mutation, then selection, in two segregated phases. The first stage is proliferation and mutation. Starting with a selected subset of the population, each individual divides $D$ times, producing $2^D$ descendants. We assume that each division takes a fixed amount of time.  
Motivated by studies of B cell SHM \cite{Pae2025nature}, we impose that mutations occur only after the last ($D$th) division before selection. Each individual is a member of an integer fitness class $i$, and each mutation either increases or decreases fitness by one unit, which can model reasonable distributions of mutational effects \cite{GoodPNAS2012}. 
The number of mutations for each individual is drawn from a Poisson distribution with mean $\gamma$, and we assume that multiple mutations act additively. 
Each mutation has probability $p_+$ of being positive and probability $1-p_+$ of being negative, where $p_+\ll 1$ corresponds to the biologically relevant regime in which beneficial mutations are rare (Fig.~\ref{fig:1}(c)).

We define as $N$ the total population size after proliferation and mutation. We focus on the case where selection retains only the top fraction $f=1/2^D$ of these individuals ranked by fitness, and all unselected individuals are discarded. This choice of $f$  maintains a constant post-proliferation population size of $N$. 
In the Supplementary Information, we consider a generalization where a smaller top fraction $f<1/2^D$ is selected to proliferate while additional randomly chosen individuals are retained without proliferation to maintain a constant population size. The upper boundary $f=1/2^D$ achieves the highest adaptation velocities, though we note that adaptation velocities are robust over a wide range of $f$ values (see Supplementary Information).
Iterations of proliferation, mutation, and selection change the mean population fitness over time. We define the adaptation velocity as $v_{\text{adapt}}=\mathrm{d} \langle i \rangle/\mathrm{d} t$, where the fitness $i$ is averaged over the population.
We performed agent-based simulations to determine how the number of divisions $D$ before selection and the mutation rate $\gamma$ affect the adaptation velocity for fixed $N$. At $t=0$, all individuals were initialized with the same fitness, $i=0$. Averaging over 500 independent simulations, we found that the mean fitness increased approximately linearly over long times. The adaptation velocities at steady state are shown in Fig.~\ref{fig:1}(d). For each value of $D$, there is an optimal mutation rate $\gamma _\mathrm{opt}(D)$ that maximizes the adaptation velocity. We find that this optimal mutation rate scales approximately linearly with $D$, such that the average per-division optimum $\gamma _\mathrm{opt}(D)/D$ is nearly constant. Moreover, the maximum adaptation velocity achieved at this optimum is very similar for different values of $D$, albeit with a slight decrease for larger $D$. 

\begin{figure*}
    \centering
    \includegraphics[width=1\linewidth]{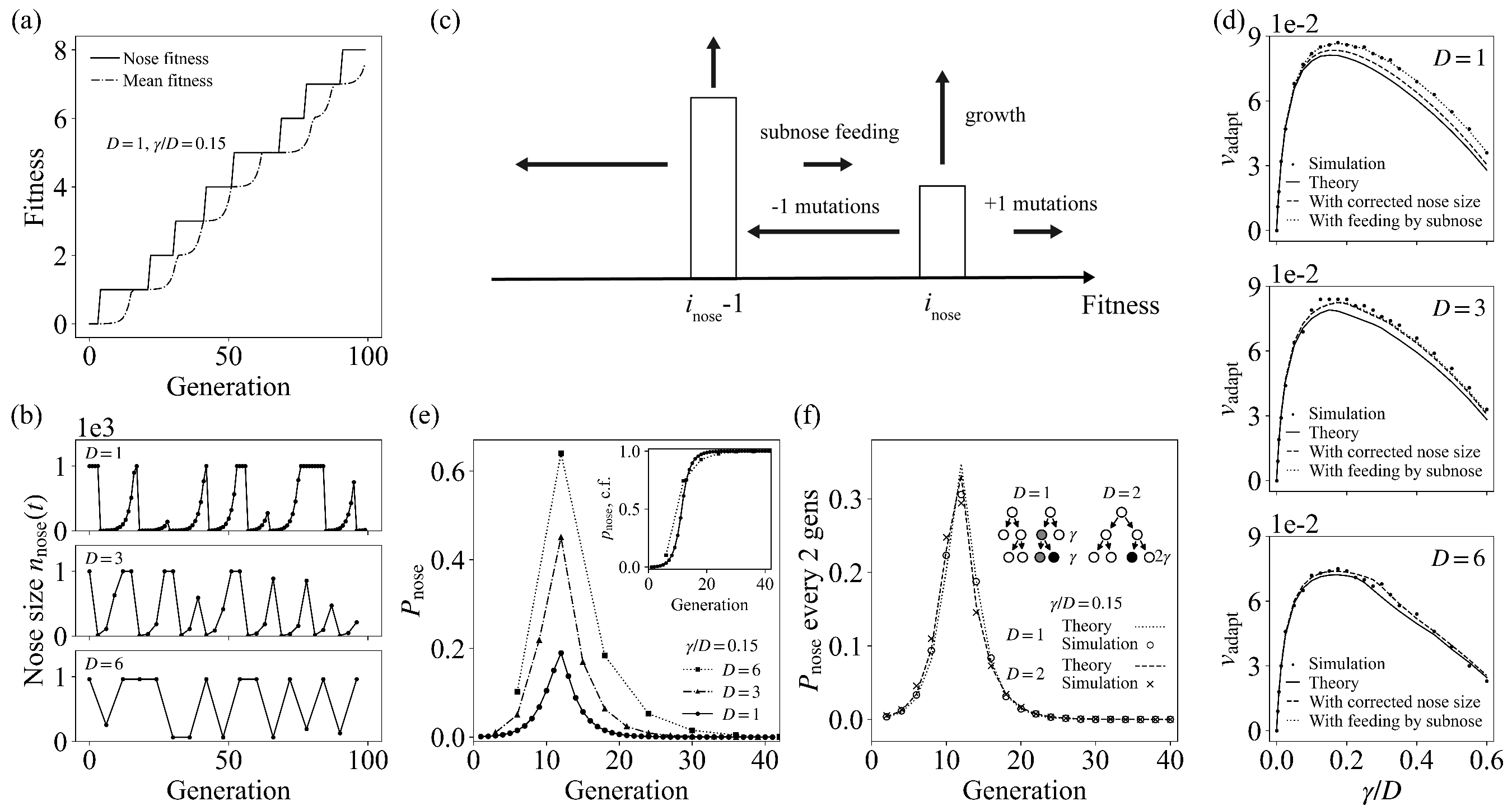}
    \caption{The dynamics of the nose predicts the adaptation velocity. (a) Fitness of the nose and of the average population over time. (b) Representative dynamics of nose size over time for $D=1,3,6$. (c) Schematics showing processes that affect nose dynamics. (d) Comparison of simulations and theory for adaptation velocity versus mutation rate scaled by $D$ for $D=1,3,6$. Unless otherwise specified, parameter values are $D=1,\gamma =0.15$. (e) Probability that a new nose emerges versus generation since previous nose formed for $D=1,3,6$ for the same $\gamma /D$, with the inset showing the cumulative probability. (f) Probability of forming a new nose after two generations, obtained from both theory and simulations, comparing $D = 1$ and $D = 2 $ cases for a fixed $\gamma /D$. Inset: Schematic comparing the cases for $D=1$ and $2$. Circle colors represent fitness phenotypes. }
    \label{fig:2}
\end{figure*}

Why are the optimal average mutation rates per division similar, and why are the corresponding adaptation velocities nearly constant? To answer these questions, we develop a theory. We note that when beneficial mutations are rare and of small incremental effect, adaptive evolution is driven by the fittest individuals in the population, often referred to as the ``nose'' \cite{desai2007genetics,Rouzine2008TheoPopBio,Brunet2008Genetics}. We therefore compared the evolution of the mean population fitness with that of the nose fitness in the same population, as shown in Fig.~\ref{fig:2}(a). The nose fitness increases in discrete steps, followed by a steady rise in the mean population fitness at the same average rate. This suggests that population-level adaptation is driven by the nose, motivating us to focus on its dynamics.

We track the nose size at the time of the last division, before mutations occur (Fig.~\ref{fig:2}(b)). For all values of $D$, the nose adaptation process consists of two stages: the generation of a new nose due to a $+1$ mutation acquired by the previous nose, followed by the growth of the new nose until it becomes large enough to generate the next new nose. For $D=1$, we see gradual increases in nose size after each division, while more substantial increases in nose size occur after every 3 divisions for $D=3$ and after every 6 divisions for $D=6$. Notably, the average period of nose growth before the next new nose formation lasts for approximately the same number of generations across all $D$ (Fig.~\ref{fig:2}(b)). This is the underlying reason why the adaptation velocities are similar for different $D$. We are then led to ask what factors determine this nose growth period.

The evolution of nose size is shaped by the combined effects of division, deleterious mutations, and beneficial mutations, as shown in Fig.~\ref{fig:2}(c). After $D$ divisions, each individual in the nose produces $2^D$ descendants. However deleterious mutations, which reduce fitness by $-1$, reduce the nose size. Together, these effects lead to exponential growth of the nose population by an average factor of approximately $2^D e^{-\gamma(1-p_+)}$ every $D$ generations, where $e^{-\gamma(1-p_+)}$ is the probability that an individual does not acquire a negative mutation (see Supplementary Information). 

After each round of divisions, assuming beneficial mutations are rare, each individual in the nose has probability $p^{(+1)}\approx p_+\gamma e^{-\gamma}$ of experiencing a beneficial mutation and no deleterious mutations, where $\gamma e^{-\gamma }$ is the probability that the individual gets one mutation. As the nose grows, the probability that at least one individual generates an offspring with a $+1$ mutation increases, eventually leading to the formation of a new nose. Then the nose growth process begins again.

Starting from the time at which a new nose is generated, we estimate the expected number of generations required for the next new nose to emerge. We assume that a newly generated nose initially consists of a single individual. During each cycle of nose growth, the nose size at the $g$th division following its formation, but before the last mutations occur, is
\begin{equation}
n_\text{nose}(g)=2^{g}\exp\left[-\gamma (1-p_+) \left (\frac{g}{D}-1 \right) \right], \label{eq1}
\end{equation}
where $g  = D, \, 2D, \,3D, \, \dots$ is an integer multiple of $D$. 
For all cases, the nose can grow until it reaches the total population size $N$, at which point its size is capped. 

Given $n_{\text{nose}}(g)$, we can calculate the probability $P_{+1}(g)$ that the current nose generates a new nose by virtue of at least one individual increasing its fitness by one:
\begin{equation}
P_{\text{+1}}(g)=1-(1-p^{(+1)})^{n_{\text{nose}}(g)} . \label{eq2}
\end{equation}
The probability $P_\mathrm{nose}(g)$ that a new nose first appears at generation $g$ must also include the probability that no new nose was generated in previous generations:
\begin{equation}
P_\text{nose}(g)=P_{\text{+1}}(g)\prod_{g'<g}(1-P_{\text{+1}}(g')) .
\end{equation}
The adaptation velocity can then be expressed in terms of the expected waiting time in generations, for generating a new nose:
\begin{align}
v_{\text{adapt}}=1/\left <g_{\text{nose}}\right > \quad \mathrm{\, with} \quad \left <g_{\text{nose}}\right >=\sum^{\infty}_{g=1} g P_{\text{nose}}(g), \quad \label{eq4}
\end{align}
where $g_\mathrm{nose}$ is the generation at which the new nose forms. Combining Eqs.~(\ref{eq1})-(\ref{eq4}), we obtain the first theory curves (solid) in Fig.~\ref{fig:2}(d).  

To compare the adaptation velocities across different numbers of divisions $D$, Fig.~\ref{fig:2}(e) shows the nose formation probability $P_\text{nose}(g)$ for $D=1,3,6$ for a fixed $\gamma /D$. These probability distributions peak at similar generations, confirming that the rate of nose formation is comparable across these cases.

Why do the $P_\mathrm{nose}(g)$ distributions peak at similar generations and result in similar nose formation rates? As an instructive example, we compare nose formation probabilities every two generations for both $D=1$ and $D=2$ (see Fig.~\ref{fig:2}(f) inset). For both $D=1$ and mutation rate $\gamma _1$, and $D=2$ and mutation rate $2\gamma _1$, the corresponding probabilities are identical (see Supplementary Information):
\begin{equation}
P_{\text{nose2gen},\gamma _1,D=1}=P_{\text{nose2gen},2\gamma _1,D=2}=2p_{+}\gamma _1
e^{-2\gamma _1}.
\end{equation}
Generalizing this comparison to the population level, Fig.~\ref{fig:2}(f) shows the simulated probability that the next nose is generated out of a previous nose, in bins of two generations, starting from the formation of the previous nose. Fig.~\ref{fig:2}(f) also shows analytical probability curves, which take into account the finite size of the previous nose (see Supplementary Information). The probability of nose formation over each two-generation interval is similar for the two cases, indicating that the adaptation dynamics are governed primarily by the total mutational input over the interval rather than by the precise timing of mutations within it.

Although the adaptation velocities are similar for all $D$, there is a small but systematic decrease in the optimal adaptation velocity between $D=1$ and $D=6$ (Fig.~\ref{fig:1}(d) and Fig.~\ref{fig:2}(d)). To understand this difference, we compare the cumulative nose generation probabilities, defined as the probability that the nose has been generated by generation $g$ since the time a previous nose is formed, for $D=1$ and $D=6$ at a fixed mutation rate per division $\gamma /D=0.15$ (Fig.~\ref{fig:2}(e), inset). Although $D=6$ is more likely than $D=1$ to generate a nose at early generations, it is also more likely to generate a nose at later generations, increasing the mean time required for nose generation. One can understand this difference as follows. As seen in Eq.~(\ref{eq2}), the probability of nose generation depends on the size of the previous nose: a larger nose size $n_\mathrm{nose}$ increases the probability that a new nose will form. Immediately after a new nose forms, the $D=6$ case has a larger nose size than the $D=1$ case (Eq.~(\ref{eq1})), leading to a higher probability of early nose generation. At later generations, however, the noses for both $D=1$ and $D=6$ have grown to saturate the population size $N$, so $D=6$ no longer has the advantage of a larger nose. Instead, because new nose formation is delayed until every six generations for $D=6$, late nose generation becomes more likely. The delay is more significant for $D=6$ than for smaller $D$, as the interval between selection events has grown to be similar to the timescale of nose generation $E(g_\mathrm{nose})$. This produces a long tail in the distribution of nose-generation times $P_\mathrm{nose}(g)$, explaining why $D=6$ has a lower adaptation velocity.

We compare the analytically predicted adaptation rate, obtained from the above analysis, with simulation results in Fig.~\ref{fig:2}(d). The analytical model correctly predicts that the optimal mutation rates at different $D$ are similar, with optimal mutation rates $\gamma_{\text{opt}}/D\approx 0.2$.

The small deviations between the theory and simulations can be corrected by considering higher-order effects. First, a newly generated nose may contain more than one individual, because multiple individuals in the existing nose may simultaneously acquire a $+1$ mutation. Second, individuals in the second-highest fitness class, which we refer to as the ``sub-nose'', can also acquire a $+1$ mutation and thereby contribute to the nose population. With these corrections (see Supplementary Information), the theoretical predictions agree well with simulations for all values of $D$ (Fig.~\ref{fig:2}(d)). The sub-nose correction is particularly important for $D=1$, where the nose size remains small for several selection cycles and thus susceptible to sub-nose influence. Third, the nose can be lost if all individuals in the nose acquire deleterious mutations; however, we find this last effect to be small.

\begin{figure}
    \centering
    \includegraphics[width=0.48\textwidth]{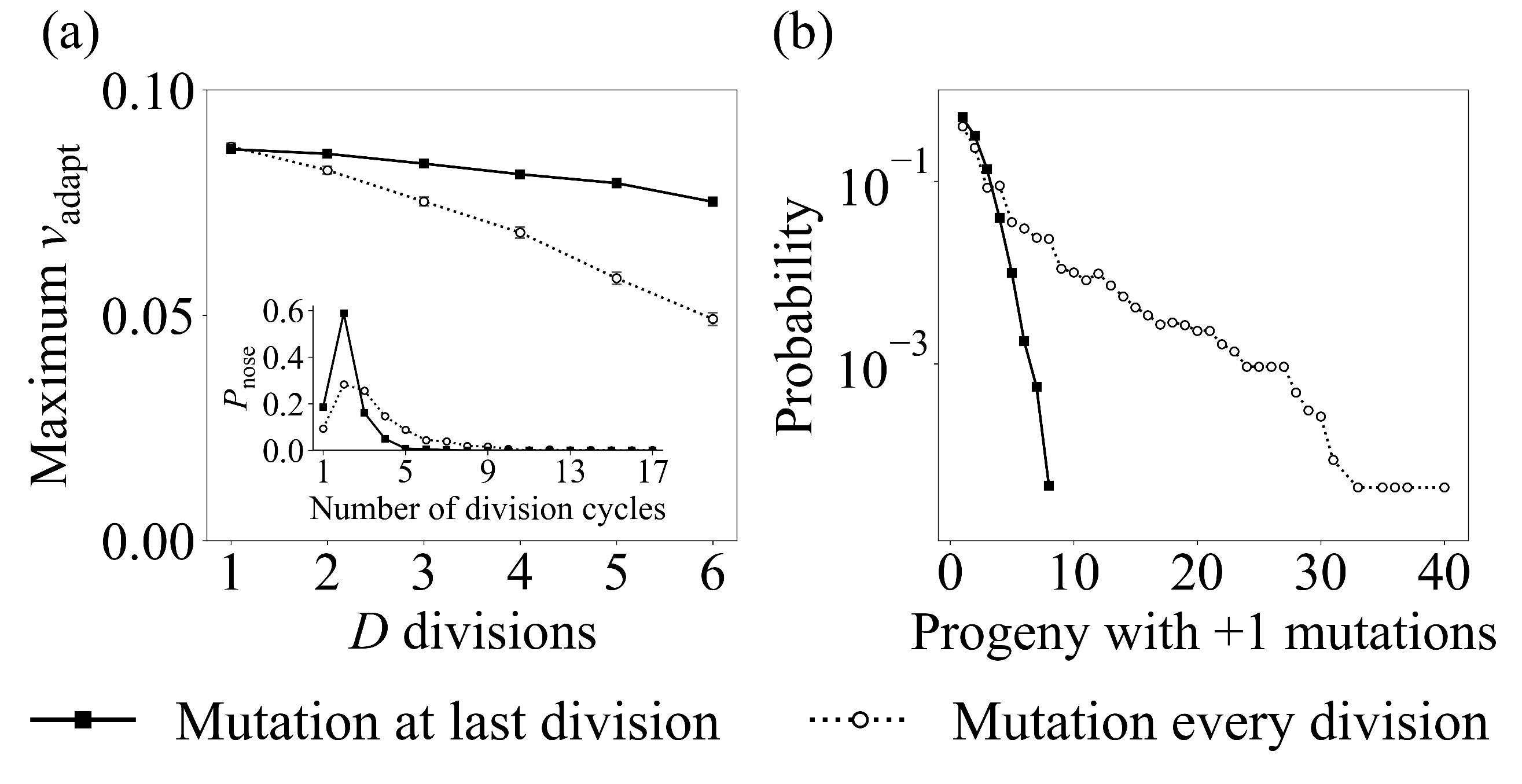}
    \caption{Effects of mutation timing on adaptation velocity and lineages. (a) Maximum adaptation velocities versus number of divisions between selection events for two different mutational protocols. Inset: For $D = 6$, waiting times for $+1$ increases in the fitness of the nose. (b) For $D=6$ and both mutational protocols, distribution of the number of final progeny with $+1$ mutations (and no $-1$ mutations) obtained in the population over a cycle of $D$ divisions, provided that a new nose is formed out of that cycle.  In both the inset of (a) and in (b) the mutation rate per division is $\gamma /D=0.15$.}
    \label{fig:S1}
\end{figure}

While we have focused on the case where mutations occur only at the last of every $D$ divisions, we now compare with the alternative case in which mutations occur at every division. We find that the optimal mutation rate per division $\gamma_\mathrm{opt}/D$, is similar across $D$ for both cases (Fig.~S5(b)), still explainable by the logic of Eq.~(5). However, the optimal adaptation velocity decreases faster with $D$ if mutations occur every division (Fig.~\ref{fig:S1}(a)). We explain this difference in adaptation rates in terms of lineages: with mutations occurring every division, early beneficial mutations will be inherited by multiple descendants before selection, producing a Luria--Delbruck-like amplification of clone size as seen in Fig.~\ref{fig:S1}(b), and in Fig.~S5(c) for a single progenitor. As by construction the average number of mutations acquired by individuals is conserved between the two mutational protocols, the more clonal case has a lower per-cycle probability of yielding new $+1$ mutations. Since at least one new $+1$ mutation is required for the formation of a new nose, the case with mutations occurring every division results in slower generation of new noses (Fig.~\ref{fig:S1}(a) inset, see also Supplementary Information) and hence a slower overall adaptation velocity. This observation may provide insight into why in B cell affinity maturation mutations only occur at the final dark-zone division \cite{Pae2025nature}, as this timing can potentially accelerate adaptation.

In this paper, motivated by the evolution of B cells in germinal centers, we built a minimal model to study how adaptation velocity depends on both frequency of selection and mutation rate. Using numerical simulations and analytical calculations, we showed that in the typical biological case where beneficial mutations are rare and of small effect, the optimal adaptation velocity is approximately invariant across selection frequencies, and that the optimal mutation rate scales with the number of divisions before selection. We accounted for this behavior by analyzing the dynamics of the ``nose'' of the fitness distribution, which determines adaptation velocity.

Our results depend on several simplifying assumptions. First, we assign a time cost only to B cell divisions. In the germinal center, the selection process involves antigen capture, presentation, interactions with Tfh cells, and migrations between dark and light zones \cite{shulman2014dynamic,BeltmanPNAS2011}, which impose an additional time cost on frequent selection, thus favoring multiple divisions in the dark zone. Second, we model selection as deterministic and perfectly rank ordered, whereas in reality selection is mediated by stochastic antigen extraction and cell-cell interactions \cite{TasScience2016,JiangPNAS2024,XuPRXLife2026}. The model also omits several features that may shape affinity maturation: Some B cell receptor mutations can have non-additive effects on affinity \cite{AdamsCellSystems2019}; B cell clones can compete for distinct epitopes \cite{KuraokaImmunity2016,SilvaCellReports2017}; and germinal centers have an additional role of maintaining B cell repertoire diversity \cite{HagglofCell2023,Scheibner2026arXiv}. To make the model analytically tractable, we make several other simplifying assumptions that we do not expect to qualitatively alter our conclusions: all cells divide and undergo selection synchronously; the generation time is fixed with no stochasticity; mutational effects on fitness are restricted to $+1$ or $-1$ while neutral mutations are neglected. Despite its simplicity, we believe the minimal model provides a useful conceptual starting point that can be readily extended to incorporate additional realistic factors.

One noteworthy point is that germinal center B cells undergo variable numbers of divisions before selection, with B cells with higher affinity receptors undergoing more divisions \cite{Gitlin2014nature}. In addition, as in our model, SHM occurs only at the final division, which, when coupled to variable numbers of divisions, has been argued to help high-affinity lineages avoid deleterious mutations while allowing less-fit lineages to explore phenotypes more broadly \cite{Pae2025nature, Merkenschlagernature2025}. The germinal center's ability to assign B cells variable numbers of divisions potentially increases the rate of adaptation beyond that attainable with uniform division numbers and mutation rates. This will be a fruitful subject for future study.

Our periodically selecting system has relevance beyond B cells. In natural systems such as host-pathogen interactions, selection can be periodic, with the frequency varying with host immunity. Experimental evolution studies can also directly control selection intervals and bottlenecks, and have found that mutation rates evolve under these conditions \cite{Sniegowski1997,Wei2022}. Our results suggest testing in a microbial setting, selecting through methods such as Fluorescence-Activated Cell Sorting, whether evolved mutation rates per division and adaptation velocities are similar across selection intervals spanning different numbers of microbial generations. More broadly, while the response to periodic environments is often studied through the lens of phenotypic switching, recombination, migration, and niche breadth \cite{KussellLeibler2005,Liberman2011,Carja2014,GerlandPNAS2009,SachdevaPNAS2020}, our framework highlights mutation-rate control as an additional strategy for adapting to a periodically selective environment. 

\textit{Acknowledgement} --- The authors would like to thank Haochen Fu for a careful reading of the manuscript. CS acknowledges support from Princeton University through the Princeton Center for Theoretical Science. This work was supported in part by Princeton University through the Center for the Physics of Biological Function. Research reported in this publication was supported by the National Institute of General Medical Sciences (NIGMS) of the National Institutes of Health (NIH) under award number R35GM156427. The content is solely the responsibility of the authors and does not necessarily represent the official views of the National Institutes of Health.

\bibliography{references}

@article{Gitlin2014nature,
  author = {Gitlin, Alexander D. and Shulman, Ziv and Nussenzweig, Michel C.},
  title = {Clonal selection in the germinal centre by regulated proliferation and hypermutation},
  journal = {Nature},
  year = {2014},
  volume = {509},
  number = {7502},
  pages = {637--640},
  doi = {10.1038/nature13300},
  url = {http://dx.doi.org/10.1038/nature13300}
}

@article{victora2012,
   author = "Victora, Gabriel D. and Nussenzweig, Michel C.",
   title = "Germinal Centers", 
   journal= "Annu. Rev. Immunol.",
   year = "2012",
   volume = "30",
   number = "Volume 30, 2012",
   pages = "429-457",
   doi = "https://doi.org/10.1146/annurev-immunol-020711-075032",
   url = "https://www.annualreviews.org/content/journals/10.1146/annurev-immunol-020711-075032",
   publisher = "Annual Reviews",
   issn = "1545-3278",
   type = "Journal Article",
  }

@ARTICLE{Merkenschlagernature2025,
  title     = "Regulated somatic hypermutation enhances antibody affinity
               maturation",
  author    = "Merkenschlager, Julia and Pyo, Andrew G T and Silva Santos,
               Gabriela S and Schaefer-Babajew, Dennis and Cipolla, Melissa and
               Hartweger, Harald and Gitlin, Alexander D and Wingreen, Ned S
               and Nussenzweig, Michel C",
  journal   = "Nature",
  publisher = "Springer Science and Business Media LLC",
  volume    =  641,
  number    =  8062,
  pages     = "495--502",
  month     =  may,
  year      =  2025,
  copyright = "https://creativecommons.org/licenses/by/4.0",
}

@ARTICLE{Pae2025nature,
  title    = "Transient silencing of hypermutation preserves {B} cell affinity
              during clonal bursting",
  author   = "Pae, Juhee and Schwan, Niklas and Ottino-Loffler, Bertrand and
              DeWitt, William S and Garg, Amar and Bortolatto, Juliana and
              Vora, Ashni A and Shen, Jin-Jie and Hobbs, Alvaro and Castro,
              Tiago B R and Mesin, Luka and Matsen, 4th, Frederick A and
              Meyer-Hermann, Michael and Victora, Gabriel D",
  journal  = "Nature",
  volume   =  641,
  number   =  8062,
  pages    = "486--494",
  month    =  may,
  year     =  2025}

@article{desai2007genetics,
  title   = {Beneficial Mutation--Selection Balance and the Effect of Linkage on Positive Selection},
  author  = {Desai, Michael M. and Fisher, Daniel S.},
  journal = {Genetics},
  volume  = {176},
  number  = {3},
  pages   = {1759--1798},
  year    = {2007},
  doi     = {10.1534/genetics.106.067678}
}

@article{shulman2014dynamic,
  author  = {Shulman, Ziv and Gitlin, Alexander D. and Weinstein, Jason S. and Lainez, Bego{\~n}a and Esplugues, Enric and Flavell, Richard A. and Craft, Joseph E. and Nussenzweig, Michel C.},
  title   = {Dynamic signaling by {T} follicular helper cells during germinal center {B} cell selection},
  journal = {Science},
  year    = {2014},
  volume  = {345},
  number  = {6200},
  pages   = {1058--1062},
  doi     = {10.1126/science.1257861}
}

@article{Sniegowski1997,
author  = {Sniegowski, Paul D. and Gerrish, Philip J. and Lenski, Richard E.},
title   = {Evolution of High Mutation Rates in Experimental Populations of {E. coli}},
journal = {Nature},
volume  = {387},
number  = {6634},
pages   = {703--705},
year    = {1997},
doi     = {10.1038/42701}
}

@article{Wei2022,
author  = {Wei, Wen and Ho, Wei-Chin and Behringer, Megan G. and Miller, Samuel F. and Bcharah, George and Lynch, Michael},
title   = {Rapid Evolution of Mutation Rate and Spectrum in Response to Environmental and Population-Genetic Challenges},
journal = {Nat. Commun.},
volume  = {13},
pages   = {4752},
year    = {2022},
doi     = {10.1038/s41467-022-32353-6}
}

@article{KussellLeibler2005,
author  = {Kussell, Edo and Leibler, Stanislas},
title   = {Phenotypic Diversity, Population Growth, and Information in Fluctuating Environments},
journal = {Science},
volume  = {309},
number  = {5743},
pages   = {2075--2078},
year    = {2005},
doi     = {10.1126/science.1114383}
}

@article{Liberman2011,
author  = {Liberman, Uri and Van Cleve, Jeremy and Feldman, Marcus W.},
title   = {On the Evolution of Mutation in Changing Environments: Recombination and Phenotypic Switching},
journal = {Genetics},
volume  = {187},
number  = {3},
pages   = {837--851},
year    = {2011},
doi     = {10.1534/genetics.110.123620}
}

@article{Carja2014,
author  = {Carja, Oana and Liberman, Uri and Feldman, Marcus W.},
title   = {Evolution in Changing Environments: Modifiers of Mutation, Recombination, and Migration},
journal = {Proc. Natl. Acad. Sci. U.S.A.},
volume  = {111},
number  = {50},
pages   = {17935--17940},
year    = {2014},
doi     = {10.1073/pnas.1417664111}
}

@article{PyoPRL2025,
  title = {Evolutionary Benefits of Fitness-Dependent Mutation Rates},
  author = {Pyo, Andrew G. T. and Yu, Qiwei and Merkenschlager, Julia and Wingreen, Ned S.},
  journal = {Phys. Rev. Lett.},
  volume = {135},
  issue = {4},
  pages = {048402},
  numpages = {6},
  year = {2025},
  month = {Jul},
  publisher = {American Physical Society},
  doi = {10.1103/szpk-sx1c},
  url = {https://link.aps.org/doi/10.1103/szpk-sx1c}
}

@article{JiangPNAS2024,
author = {Hongda Jiang  and Shenshen Wang },
title = {Physical extraction of antigen and information},
journal = {Proc. Natl. Acad. Sci. U.S.A.},
volume = {121},
number = {39},
pages = {e2320537121},
year = {2024},
doi = {10.1073/pnas.2320537121},
URL = {https://www.pnas.org/doi/abs/10.1073/pnas.2320537121}}

@article {Scheibner2026arXiv,
	author = {Scheibner, Colin and Pyo, Andrew G.T. and Schaefer-Babajew, Dennis and Wingreen, Ned S.},
	title = {Antigen-tethering proteins on follicular dendritic cells influence the affinity and diversity of germinal-center B cells},
	elocation-id = {2026.01.14.699581},
	year = {2026},
	publisher = {Cold Spring Harbor Laboratory},
	URL = {https://www.biorxiv.org/content/early/2026/02/24/2026.01.14.699581},
	journal = {bioRxiv}
}

@article{Allen2007Science,
author = {Christopher D. C. Allen  and Takaharu Okada  and H. Lucy Tang  and Jason G. Cyster },
title = {Imaging of Germinal Center Selection Events During Affinity Maturation},
journal = {Science},
volume = {315},
number = {5811},
pages = {528-531},
year = {2007},
doi = {10.1126/science.1136736},
URL = {https://www.science.org/doi/abs/10.1126/science.1136736}}

@article{HagglofCell2023,
title = {Continuous germinal center invasion contributes to the diversity of the immune response},
journal = {Cell},
volume = {186},
number = {1},
pages = {147-161.e15},
year = {2023},
issn = {0092-8674},
doi = {https://doi.org/10.1016/j.cell.2022.11.032},
url = {https://www.sciencedirect.com/science/article/pii/S0092867422015070},
author = {Thomas Hägglöf and Melissa Cipolla and Maximilian Loewe and Spencer T. Chen and Ervin E. Kara and Luka Mesin and Harald Hartweger and Mohamed A. ElTanbouly and Alice Cho and Anna Gazumyan and Victor Ramos and Leonidas Stamatatos and Thiago Y. Oliveira and Michel C. Nussenzweig and Charlotte Viant}
}

@article{BeltmanPNAS2011,
author = {Joost B. Beltman  and Christopher D. C. Allen  and Jason G. Cyster  and Rob J. de Boer },
title = {B cells within germinal centers migrate preferentially from dark to light zone},
journal = {Proc. Natl. Acad. Sci. U.S.A.},
volume = {108},
number = {21},
pages = {8755-8760},
year = {2011},
doi = {10.1073/pnas.1101554108},
URL = {https://www.pnas.org/doi/abs/10.1073/pnas.1101554108}}

@article{XuPRXLife2026,
  title = {Adaptive Mechanical Proofreading Toward Robust yet Tunable Humoral Immunity},
  author = {Xu, Qing and Jiang, Hongda and Wang, Shenshen},
  journal = {PRX Life},
  volume = {4},
  issue = {2},
  pages = {023007},
  numpages = {12},
  year = {2026},
  month = {Apr},
  publisher = {American Physical Society},
  doi = {10.1103/vfqt-t3b9},
  url = {https://link.aps.org/doi/10.1103/vfqt-t3b9}
}

@article{Rouzine2008TheoPopBio,
title = {The traveling-wave approach to asexual evolution: Muller's ratchet and speed of adaptation},
journal = {Theor. Popul. Biol.},
volume = {73},
number = {1},
pages = {24-46},
year = {2008},
issn = {0040-5809},
doi = {https://doi.org/10.1016/j.tpb.2007.10.004},
url = {https://www.sciencedirect.com/science/article/pii/S0040580907001207},
author = {Igor M. Rouzine and Éric Brunet and Claus O. Wilke}
}

@article{Brunet2008Genetics,
    author = {Brunet, Éric and Rouzine, Igor M and Wilke, Claus O},
    title = {The Stochastic Edge in Adaptive Evolution},
    journal = {Genetics},
    volume = {179},
    number = {1},
    pages = {603-620},
    year = {2008},
    month = {05},
    issn = {1943-2631},
    doi = {10.1534/genetics.107.079319},
    url = {https://doi.org/10.1534/genetics.107.079319}}

@Article{PelissierCells2020,
AUTHOR = {Pélissier, Aurélien and Akrout, Youcef and Jahn, Katharina and Kuipers, Jack and Klein, Ulf and Beerenwinkel, Niko and Rodríguez Martínez, María},
TITLE = {Computational Model Reveals a Stochastic Mechanism behind Germinal Center Clonal Bursts},
JOURNAL = {Cells},
VOLUME = {9},
YEAR = {2020},
NUMBER = {6},
ARTICLE-NUMBER = {1448},
URL = {https://www.mdpi.com/2073-4409/9/6/1448},
PubMedID = {32532145},
ISSN = {2073-4409},
DOI = {10.3390/cells9061448}
}

@article{Wittenbrinkjimmun2010,
    author = {Wittenbrink, Nicole and Weber, Tom S and Klein, Anke and Weiser, Armin A and Zuschratter, Werner and Sibila, Michael and Schuchhardt, Johannes and Or-Guil, Michal},
    title = {Broad Volume Distributions Indicate Nonsynchronized Growth and Suggest Sudden Collapses of Germinal Center B Cell Populations},
    journal = {J. Immunol.},
    volume = {184},
    number = {3},
    pages = {1339-1347},
    year = {2010},
    month = {02},
    issn = {0022-1767},
    doi = {10.4049/jimmunol.0901040}
}

@article{TasScience2016,
  author  = {Tas, Jeroen M. J. and Mesin, Luka and Pasqual, Giulia
             and Targ, Sasha and Jacobsen, Johanne T. and Mano, Yasuko M.
             and Chen, Casie S. and Weill, Jean-Claude
             and Reynaud, Claude-Agn{\`e}s and Browne, Edward P.
             and Meyer-Hermann, Michael and Victora, Gabriel D.},
  title   = {Visualizing Antibody Affinity Maturation in Germinal Centers},
  journal = {Science},
  year    = {2016},
  volume  = {351},
  number  = {6277},
  pages   = {1048--1054},
  doi     = {10.1126/science.aad3439}
}

@article{AdamsCellSystems2019,
  author  = {Adams, Rhys M. and Kinney, Justin B.
             and Walczak, Aleksandra M. and Mora, Thierry},
  title   = {Epistasis in a Fitness Landscape Defined by
             Antibody--Antigen Binding Free Energy},
  journal = {Cell Syst.},
  year    = {2019},
  volume  = {8},
  number  = {1},
  pages   = {86--93.e3},
  doi     = {10.1016/j.cels.2018.12.004}
}

@article{KuraokaImmunity2016,
  author  = {Kuraoka, Masayuki and Schmidt, Aaron G. and Nojima, Takuya
             and Feng, Feng and Watanabe, Akiko and Kitamura, Daisuke
             and Harrison, Stephen C. and Kepler, Thomas B.
             and Kelsoe, Garnett},
  title   = {Complex Antigens Drive Permissive Clonal Selection
             in Germinal Centers},
  journal = {Immunity},
  year    = {2016},
  volume  = {44},
  number  = {3},
  pages   = {542--552},
  doi     = {10.1016/j.immuni.2016.02.010}
}

@article{SilvaCellReports2017,
  author  = {Silva, Murillo and Nguyen, Thao H. and Philbrook, Phaethon
             and Chu, Matthew and Sears, Olivia and Hatfield, Stephen
             and Abbott, Robert K. and Kelsoe, Garnett
             and Sitkovsky, Michail V.},
  title   = {Targeted Elimination of Immunodominant B Cells Drives
             the Germinal Center Reaction toward Subdominant Epitopes},
  journal = {Cell Rep.},
  year    = {2017},
  volume  = {21},
  number  = {13},
  pages   = {3672--3680},
  doi     = {10.1016/j.celrep.2017.12.014}}

@article{
GerlandPNAS2009,
author = {Ulrich Gerland  and Terence Hwa },
title = {Evolutionary selection between alternative modes of gene regulation},
journal = {Proc. Natl. Acad. Sci. U.S.A.},
volume = {106},
number = {22},
pages = {8841-8846},
year = {2009},
doi = {10.1073/pnas.0808500106},
URL = {https://www.pnas.org/doi/abs/10.1073/pnas.0808500106}}

@article{GoodPNAS2012,
author = {Benjamin H. Good  and Igor M. Rouzine  and Daniel J. Balick  and Oskar Hallatschek  and Michael M. Desai },
title = {Distribution of fixed beneficial mutations and the rate of adaptation in asexual populations},
journal = {Proc. Natl. Acad. Sci. U.S.A.},
volume = {109},
number = {13},
pages = {4950-4955},
year = {2012},
doi = {10.1073/pnas.1119910109},
URL = {https://www.pnas.org/doi/abs/10.1073/pnas.1119910109}}

@article{WardlawGenetics2012,
    author = {Wardlaw, Alison M and Agrawal, Aneil F},
    title = {Temporal Variation in Selection Accelerates Mutational Decay by Muller’s Ratchet},
    journal = {Genetics},
    volume = {191},
    number = {3},
    pages = {907-916},
    year = {2012},
    month = {07},
    issn = {1943-2631},
    doi = {10.1534/genetics.112.140962},
    url = {https://doi.org/10.1534/genetics.112.140962}
}

@article{TrubenovaEvolution2019,
author = {Trubenová, Barbora and Krejca, Martin S. and Lehre, Per Kristian and Kötzing, Timo},
title = {Surfing on the seascape: Adaptation in a changing environment},
journal = {Evolution},
volume = {73},
number = {7},
pages = {1356-1374},
doi = {https://doi.org/10.1111/evo.13784},
url = {https://onlinelibrary.wiley.com/doi/abs/10.1111/evo.13784},
year = {2019}
}

@article{GyorgyNatureComm2023,
  author  = {Gyorgy, Andras},
  title   = {Competition and evolutionary selection among core regulatory motifs in gene expression control},
  journal = {Nat. Commun.},
  year    = {2023},
  volume  = {14},
  pages   = {8266},
  doi     = {10.1038/s41467-023-43327-7}
}

@article{NourmohammadMBE2019,
    author = {Nourmohammad, Armita and Otwinowski, Jakub and Łuksza, Marta and Mora, Thierry and Walczak, Aleksandra M},
    title = {Fierce Selection and Interference in B-Cell Repertoire Response to Chronic HIV-1},
    journal = {Mol. Biol. Evol.},
    volume = {36},
    number = {10},
    pages = {2184-2194},
    year = {2019},
    month = {10},
    issn = {0737-4038},
    doi = {10.1093/molbev/msz143}
}

@article{DespondsPNAS2016,
author = {Jonathan Desponds  and Thierry Mora  and Aleksandra M. Walczak },
title = {Fluctuating fitness shapes the clone-size distribution of immune repertoires},
journal = {Proc. Natl. Acad. Sci. U.S.A.},
volume = {113},
number = {2},
pages = {274-279},
year = {2016},
doi = {10.1073/pnas.1512977112}}

@article{
SachdevaPNAS2020,
author = {Vedant Sachdeva  and Kabir Husain  and Jiming Sheng  and Shenshen Wang  and Arvind Murugan },
title = {Tuning environmental timescales to evolve and maintain generalists},
journal = {Proc. Natl. Acad. Sci. U.S.A.},
volume = {117},
number = {23},
pages = {12693-12699},
year = {2020},
doi = {10.1073/pnas.1914586117}}

@article{DeWittCell2026,
title = {Replaying germinal center evolution on a quantified affinity landscape},
journal = {Cell},
volume = {189},
number = {16},
pages = {4980-4996.e8},
year = {2026},
issn = {0092-8674},
doi = {https://doi.org/10.1016/j.cell.2026.05.013},
author = {William S. DeWitt and Ashni A. Vora and Tatsuya Araki and Jared G. Galloway and Tanwee Alkutkar and Juliana Bortolatto and Tiago B.R. Castro and Will Dumm and Chris Jennings-Shaffer and Tongqiu Jia and Luka Mesin and Gabriel Ozorowski and Juhee Pae and Duncan K. Ralph and Jesse D. Bloom and Armita Nourmohammad and Yun S. Song and Andrew B. Ward and Tyler N. Starr and Frederick A. Matsen and Gabriel D. Victora}
}

\end{document}

\title{Supplementary Information: Optimizing Evolution Under Cyclic Selection:\\ the Case of B Cells}

\author{Xingjian Hou}
\affiliation{Lewis-Sigler Institute for Integrative Genomics, Princeton University, Princeton, New Jersey 08544, USA}

\author{Colin Scheibner\textsuperscript{$\dagger$}}
\affiliation{Joseph Henry Laboratories of Physics, Princeton University, Princeton, New Jersey 08544, USA}
\affiliation{Princeton Center for Theoretical Science, Princeton University, Princeton, New Jersey 08544, USA}

\author{Ned S. Wingreen\textsuperscript{$\dagger$}}
\affiliation{Lewis-Sigler Institute for Integrative Genomics, Princeton University, Princeton, New Jersey 08544, USA}
\affiliation{Department of Molecular Biology, Princeton University, Princeton, New Jersey 08544, USA}

\maketitle

\tableofcontents

\setcounter{figure}{0}
\renewcommand{\thefigure}{S\arabic{figure}} 

\renewcommand{\theequation}{S\arabic{equation}}

\section{I. Growth rate of the nose}

In the main text, we obtained an expression for the nose size $n_\mathrm{nose}(g)$, where $g$ is the number of generations since the current nose formed, assuming that no new nose has formed in the meantime [Eq.~(1) of the main text]. Here, we derive the nose growth factor that enters this expression.

In our model of periodic selection, each individual divides $D$ times between selection events. Mutations occur only after the last of these divisions, and the number of mutations acquired by each individual is drawn from a Poisson distribution with mean $\gamma$. Each mutation independently increases fitness by one with probability $p_+$ or decreases it by one with probability $1-p_+$, where $p_+\ll 1$.

Nose growth is set by the competing effects of division and mutation. Division increases the nose size by a factor of $2^D$ per selection cycle, whereas deleterious mutations, which reduce fitness by one, remove individuals from the nose and thereby multiply its size by $e^{-\gamma(1-p_+)}$. Until a new nose forms, the nose therefore grows by a factor of $2^De^{-\gamma(1-p_+)}$ per selection cycle, i.e., every $D$ generations. We now derive the factor $e^{-\gamma(1-p_+)}$.

This factor is the probability that an individual in the nose acquires no mutations, and hence remains in the nose, given that it does not acquire a beneficial mutation (which would instead create a new nose). Let $\mu_+$ and $\mu_-$ denote the numbers of beneficial and deleterious mutations acquired by an individual. Because each mutation is independently beneficial or deleterious, Poisson thinning gives
\begin{equation}
    \mu_+ \sim \operatorname{Poisson}(\gamma p_+),
\qquad
    \mu_- \sim \operatorname{Poisson}(\gamma(1-p_+)).
\end{equation}
Since the total number of mutations is $\mu = \mu_+ + \mu_-$, the probability of acquiring no mutations, conditioned on acquiring no beneficial mutations, is
\begin{equation}
\begin{aligned}
P(\mu = 0 \mid \mu_+ = 0)
&= P(\mu_- = 0 \mid \mu_+ = 0) \\
&= P(\mu_- = 0) \\
&= e^{-\gamma(1-p_+)},
\end{aligned}
\end{equation}
where the second equality follows from the independence of $\mu_+$ and $\mu_-$.

\section{II. An intermediate mutation rate optimizes adaptation velocity}

\begin{figure}[H]
    \centering
    \includegraphics[width=0.3\linewidth]{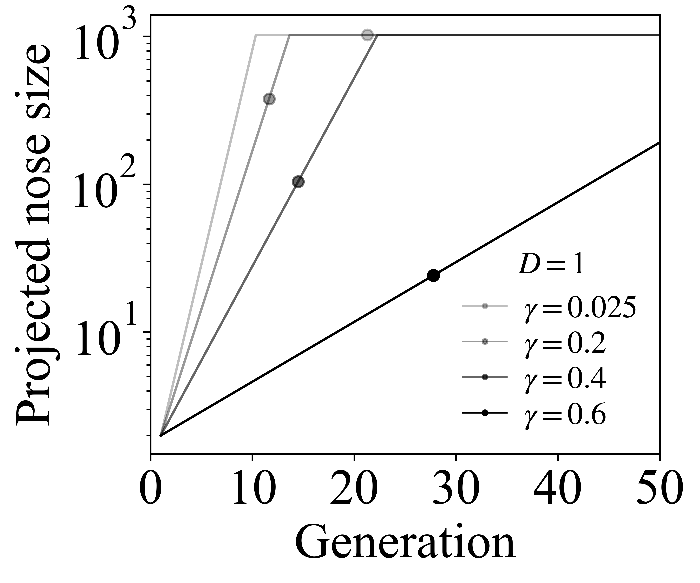}
    \captionsetup[figure]{
    justification=centering,
    singlelinecheck=false}
    \caption{Projected nose size versus generation, assuming no new nose forms (curves), and the mean generation at which a new nose forms (dots), for mutation rates $\gamma = 0.025$, $0.2$, $0.4$, and $0.6$ with $D=1$ division between selection events.}
    \label{fig:2e}
\end{figure}

In the main text, we found that the rate of new nose formation, and hence the adaptation velocity, is maximized at an intermediate mutation rate. Here, we give an intuitive explanation. For $D=1$, Fig.~\ref{fig:2e} shows the projected nose size, assuming that no new nose forms, which grows exponentially in time until it reaches the population size. When the mutation rate is very low (e.g., $\gamma=0.025$), the nose grows quickly but beneficial mutations are rare, so the waiting time for the next nose is long. When the mutation rate is very high (e.g., $\gamma=0.6$), deleterious mutations are frequent and slow the growth of the current nose. The resulting small nose generates few beneficial mutations despite the high per-individual mutation rate, which again delays formation of the next nose. An intermediate mutation rate balances these two effects and maximizes the adaptation velocity.

\section{III. Effect of selection fraction on adaptation velocity}
In the main text, we focused on the case in which the top fraction $f=1/2^D$ of individuals is selected to divide. Here, we consider selected fractions $f<1/2^D$.

\begin{figure}[H]
\centering
\includegraphics[width=0.45\linewidth]{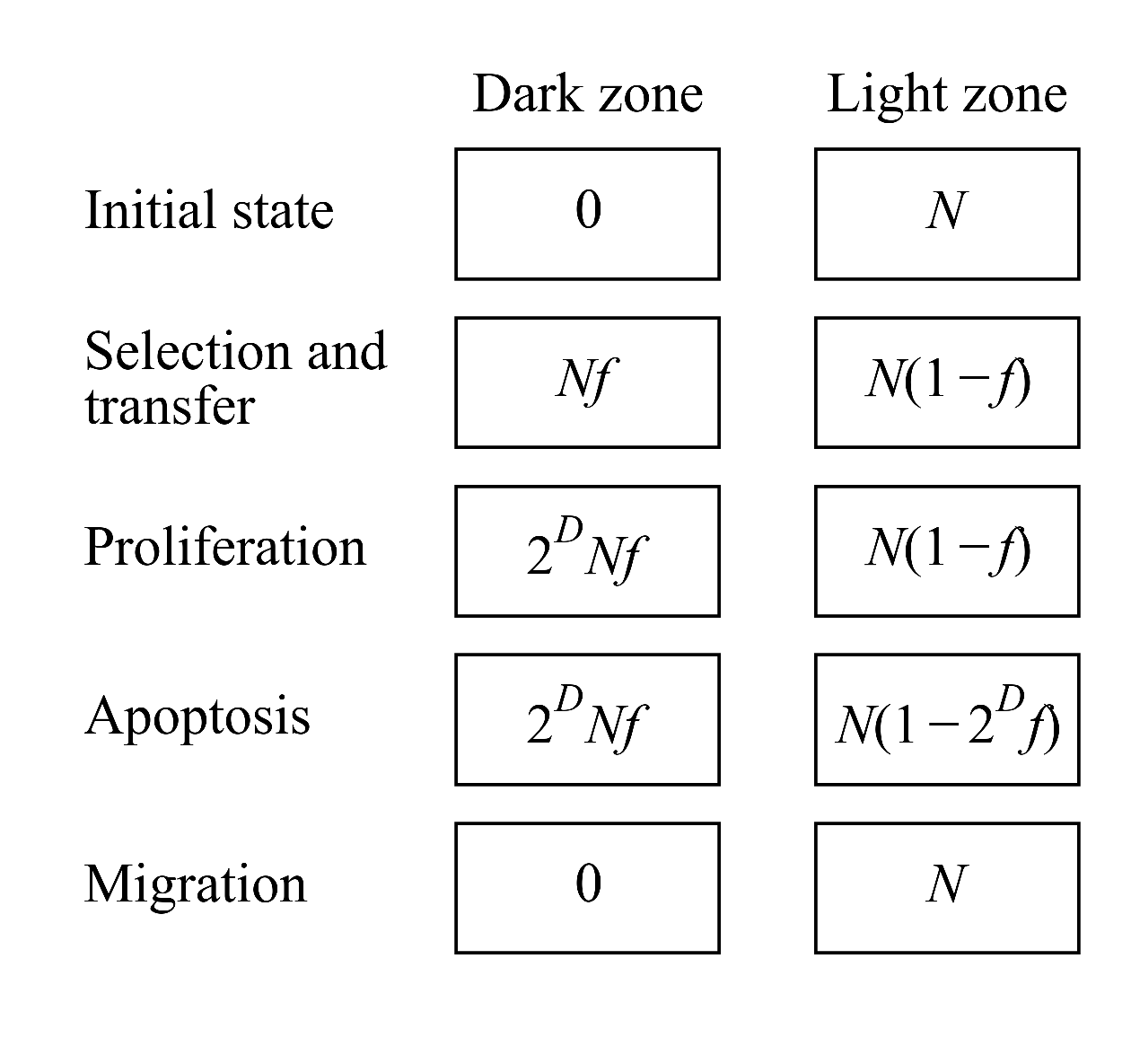}
\caption{Model of B cell population changes across the dark and light zones over one cycle of selection and proliferation.}
\label{fig}
\end{figure}

Motivated by affinity maturation, we consider a model in which B cells cycle between the light and dark zones of the germinal center (Fig.~\ref{fig}). In the light zone, the top fraction $f$ of B cells, ranked by fitness, is selected and migrates to the dark zone, where each selected cell divides $D$ times and then undergoes somatic hypermutation. Meanwhile, randomly chosen unselected B cells in the light zone undergo apoptosis, with the number of apoptotic cells chosen to keep the total population of the light and dark zones constant. The B cells that have completed proliferation in the dark zone then return to the light zone, where all cells undergo the next round of selection. We study how the selected fraction $f$, together with the number of divisions $D$ and the mutation rate $\gamma$, affects the adaptation velocity.

We first determine which values of $f$ are compatible with a constant population size. After the apoptosis step, the number of B cells remaining in the light zone is $N(1-2^Df)$, which must be non-negative. This requires
\begin{equation}
f\leq \frac{1}{2^D}, \label{stability}
\end{equation}
since otherwise the germinal center population would grow without bound. We therefore restrict attention to $f\leq 1/2^D$, and write $f_\mathrm{max}=1/2^D$.

\begin{figure}[H]
\centering
\includegraphics[width=0.9\linewidth]{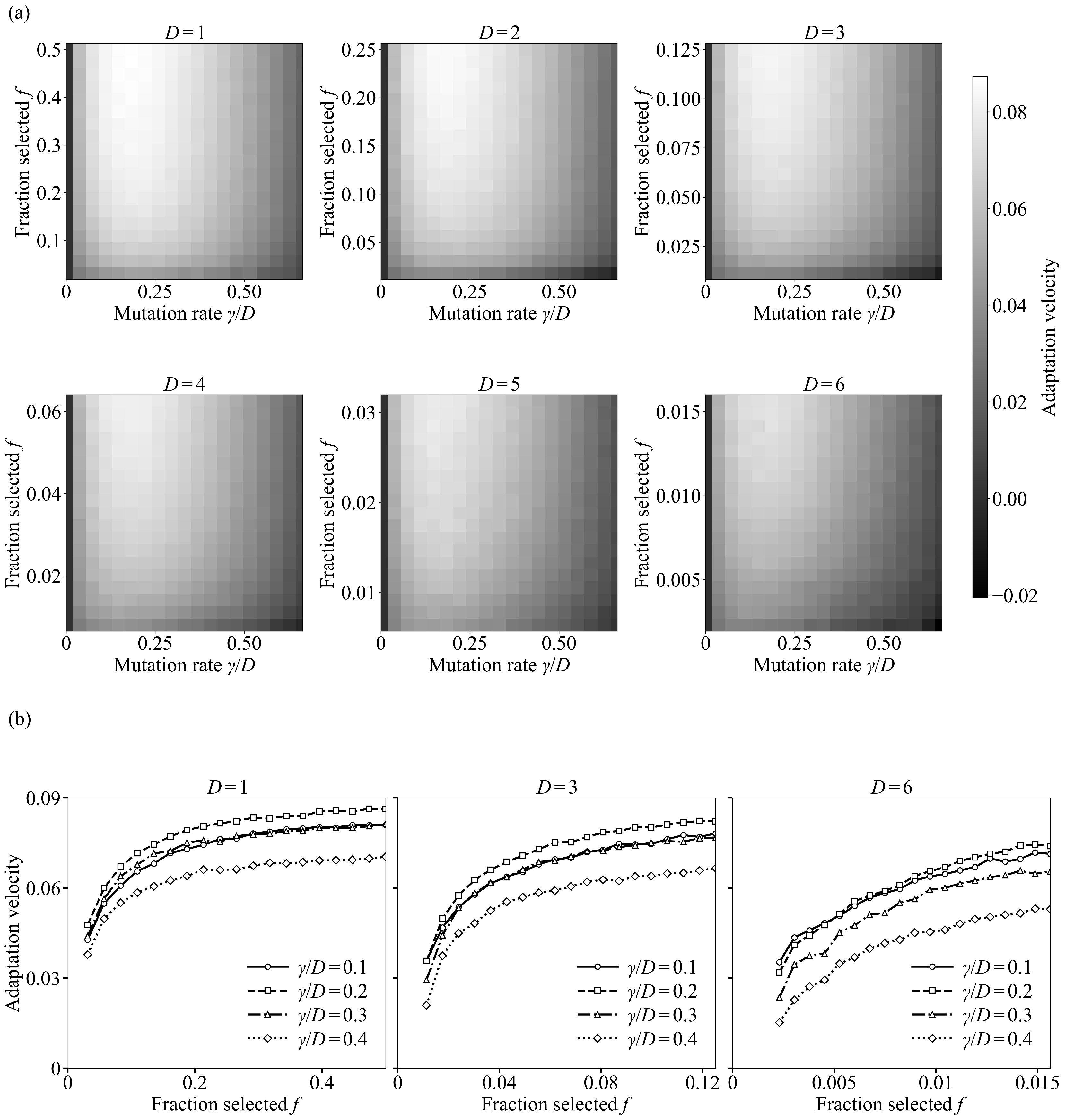}
\caption{(a) Adaptation velocity as a function of the mutation rate per division $\gamma/D$ and the selected fraction $f$, for $D=1$--$6$. (b) Adaptation velocity versus $f$ for fixed values of $\gamma/D$. In each panel, the largest value of $f$ shown is $f_\mathrm{max}=1/2^D$, the maximum compatible with a constant germinal center population.
}
\label{scanparameter}
\end{figure}

Using agent-based simulations, we computed the adaptation velocity $v_\mathrm{adapt}=\mathrm{d}\langle i\rangle/\mathrm{d}t$, averaged over 100 independent runs, as a function of the selected fraction $f$, the number of divisions $D$, and the mutation rate per division $\gamma/D$ [Fig.~\ref{scanparameter}(a)]. For all $D$, the largest allowed fraction, $f=f_\mathrm{max}$, gives the highest adaptation velocity. Because $Nf$ B cells are selected to proliferate, a larger $f$ yields more proliferating cells and hence more opportunities for beneficial mutations.

Nevertheless, the adaptation velocity varies only weakly over a wide range of $f$ and drops sharply only when $f$ is small [Fig.~\ref{scanparameter}(b)]. For example, for $D=1$ and $\gamma /D=0.2$, the adaptation velocity decreases only from 0.086 to 0.082 as $f$ is reduced from $f_\mathrm{max}=1/2$ to $f_\mathrm{max}/2=1/4$. For $D=6$, the decrease is somewhat larger, from 0.074 to 0.060 as $f$ is reduced from $f_\mathrm{max}=1/64$ to $f_\mathrm{max}/2=1/128$. We conclude that, provided $f$ is reasonably close to $f_\mathrm{max}$, the results obtained for $f=1/2^D$ are robust, with similar adaptation velocities across $D$. We explain this insensitivity in the next section.

\section{IV. Dependence of adaptation velocity on population $N$ and selection fraction $f$}
In many evolving systems, the rate of adaptation increases with population size. In our model, the number of individuals selected to proliferate is $fN$, so changing either the total population size $N$ or the selected fraction $f$ could affect the adaptation velocity.

\begin{figure}[H]
    \centering
    \includegraphics[width=0.5\linewidth]{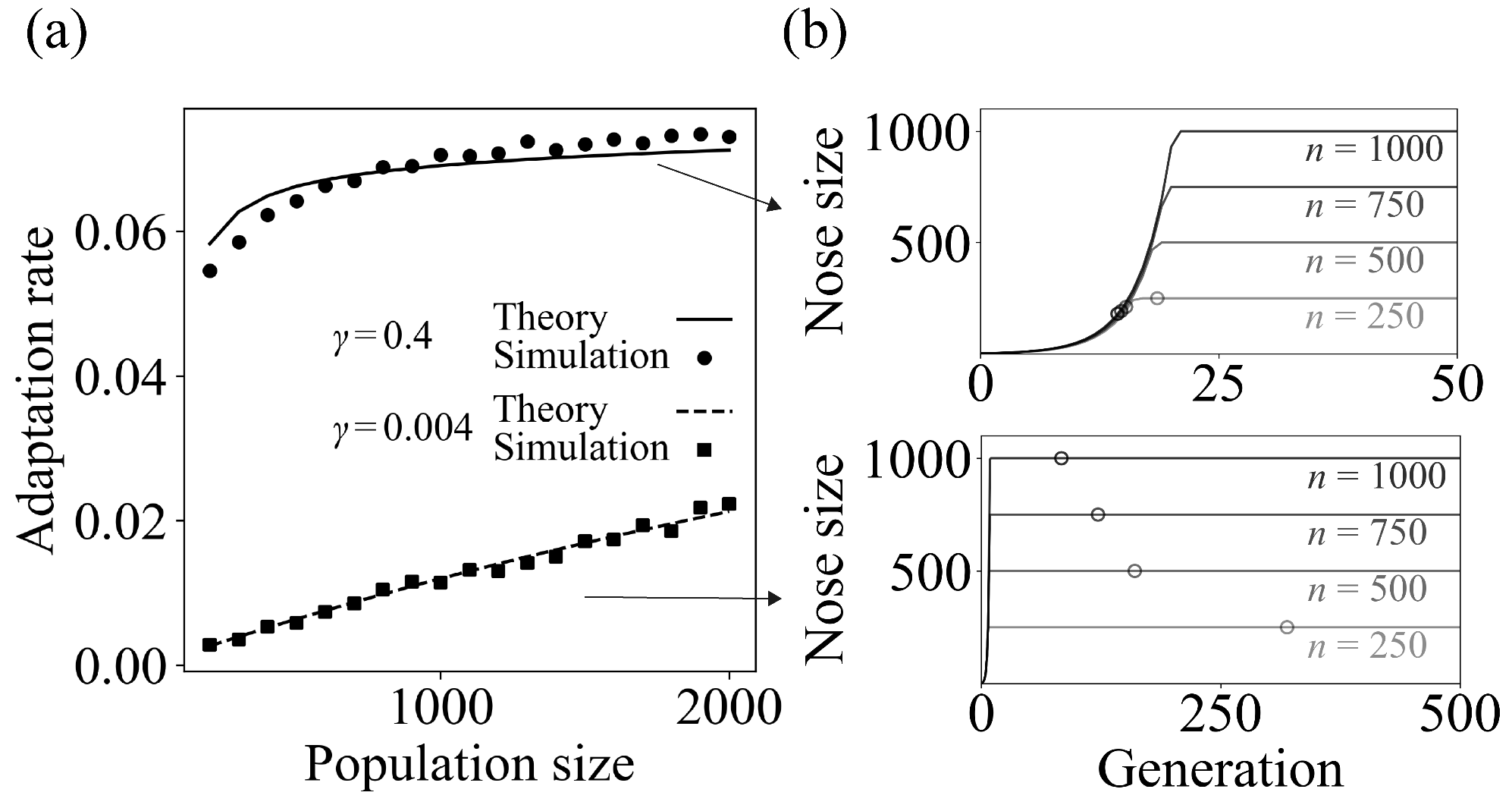}
    \caption{Effect of population size on adaptation velocity. (a) Adaptation velocity versus total population size $N$ for $D=1$, $f=1/2$, and $\gamma=0.004$ and $0.4$. (b) Curves: nose size versus generation for several $N$. Circles: nose size and generation at which a new nose forms. Upper panel: $\gamma=0.4$; lower panel: $\gamma=0.004$.}
    \label{fig:4}
\end{figure}

We first fix $f$ and vary $N$ alone. Thus far, we have used $N = 1024$, comparable to the number of B cells in a germinal center. At low mutation rates, the adaptation velocity is approximately proportional to $N$, whereas at high mutation rates it depends only weakly on $N$ [Fig.~\ref{fig:4}(a)]. At low mutation rates, beneficial mutations are rare, so many generations elapse before a new nose forms, by which time the previous nose has typically grown to the full population size $N$ [Fig.~\ref{fig:4}(b), lower panel]. In this regime, the per-generation probability of forming a new nose is $P_{+1}\approx n_\text{nose}p^{(+1)}\approx Np^{(+1)}$ [cf.~Eq.~(2) of the main text], so that $v_\text{adapt}\approx P_{+1}\propto N$. At high mutation rates, by contrast, beneficial mutations are common enough that a new nose typically forms before the previous nose reaches the cap $N$ [Fig.~\ref{fig:4}(b), upper panel]. The previous nose then grows as $n_\text{nose}(g)=2^{g}e^{-\gamma (1-p_+)(g/D-1)}$, independently of $N$, so the nose formation probability, and hence the adaptation velocity, is nearly independent of $N$.

The same argument accounts for the weak dependence of the adaptation velocity on $f$. Indeed, the high-mutation-rate curve in Fig.~\ref{fig:4}(a) has a shape similar to the curves in Fig.~\ref{scanparameter}(b): in both cases, the adaptation velocity varies little over a wide range of parameter values, because the nose rarely reaches the population cap before a new nose forms.

\section{V. Probability of nose formation over two generations}

In the main text, we compared the probability that the current nose generates a new nose over two generations for $D=1$ and $D=2$. For $D=1$, mutations occur at rate $\gamma_1$ after every division; for $D=2$, mutations occur at rate $2\gamma_1$ after every second division. Here, we derive these probabilities. Throughout, we write $p^{(+1)}(\gamma)=p_+\gamma e^{-\gamma}$ for the probability that an individual with mutation rate $\gamma$ acquires exactly one mutation and that this mutation is beneficial.

We first calculate the probability that a new nose is generated from a single lineage. For $D=1$, in the limit of rare beneficial mutations, a lineage that produces a new nose within two generations must acquire either (1) a beneficial mutation in generation 1 and no deleterious mutation in generation 2, or (2) no deleterious mutation in generation 1 and a beneficial mutation in generation 2. The probability of this event is
\begin{equation}
    P_{\text{nose2gen},\gamma_1,D=1}=2p^{(+1)}(\gamma_1)\cdot e^{-\gamma_1}=2p_+\gamma_1e^{-2\gamma_1}.
\end{equation}
where $e^{-\gamma_1}$ approximates the probability of acquiring no deleterious mutation in one generation when $p_+\ll 1$. For $D=2$, the corresponding probability is
\begin{equation}
    P_{\text{nose2gen},2\gamma_1,D=2}=p^{(+1)}(2\gamma_1)=2p_+\gamma_1e^{-2\gamma_1}.
\end{equation}
The two probabilities are therefore identical [Eq.~(5) of the main text].

At the population level, however, the size of the current nose changes from generation to generation. We now extend the calculation to account for this.

For $D=2$ with mutation rate $2\gamma_1$, the probability that a new nose forms at the end of the two generations is
\begin{equation}
Q_{\text{nose2gen},2\gamma_1,D=2}=1-(1-p^{(+1)}(2\gamma_1))^{n_\mathrm{nose,2,g=2}}.
\end{equation}
where $n_\mathrm{nose,2,g=2}$ is the nose size after the second division. For $D=1$ with mutation rate $\gamma_1$, the probability that at least one individual in the nose acquires a beneficial mutation during the two generations is
\begin{equation}
    Q_{\text{+1,2gen},\gamma_1,D=1}
    =
    1-(1-p^{(+1)}(\gamma_1))^{n_\mathrm{nose,1,g=1}+n_\mathrm{nose,1,g=2}},
\end{equation}
where $n_\mathrm{nose,1,g=1}$ and $n_\mathrm{nose,1,g=2}$ are the nose sizes in the first and second generations, respectively.

For $D=1$, however, a beneficial mutation during the two generations does not guarantee a new nose at the end of the second generation, because deleterious mutations can cancel its effect. An individual that acquires a beneficial mutation at $g=2$ must not have acquired a deleterious mutation at $g=1$; this requirement is already accounted for by $n_\mathrm{nose,1,g=2}$, which counts only individuals without deleterious mutations. An individual that acquires a beneficial mutation at $g=1$ must have at least one of its two progeny avoid a deleterious mutation at $g=2$, which occurs with probability $s=1-(1-e^{-\gamma_1})^2$.

The probability of new nose formation over two generations for $D=1$ is therefore
\begin{equation}
   Q_{\text{nose2gen},\gamma_1,D=1}
   =
    1-(1-p^{(+1)}(\gamma_1))^{n_\mathrm{nose,1,g=2}}(1-p^{(+1)}(\gamma_1)s)^{n_\mathrm{nose,1,g=1}}
   .
\end{equation}

In Fig.~2(f) of the main text, we plot the probability that a new nose first forms in each successive two-generation interval. This is the probability of nose formation during the corresponding interval, multiplied by the probability that no new nose formed in any preceding interval. Thus, we plot
\begin{equation}
Q_{\text{nose2gen, pdf}, i\gamma _1,D=i}(h)
=
Q_{\text{nose2gen},i\gamma _1,D=i}(h)
\prod_{h'<h}
\left(1-Q_{\text{nose2gen},i\gamma _1,D=i}(h')\right),
\end{equation}
where $i=1$ and $i=2$ correspond to the $D=1$ and $D=2$ cases, respectively, and $h$ indexes the successive two-generation intervals since the formation of the previous nose. Both probability distributions peak at the same generation in the plot due to similar probabilities of nose formation.

To see why these probabilities are similar, we linearize them. For $D=2$,
\begin{equation}
Q_{\text{nose2gen},2\gamma_1,D=2}\approx p^{(+1)}(2\gamma_1)n_\mathrm{nose,2,g=2}=2\gamma_1p_+e^{-2\gamma_1}n_\mathrm{nose,2,g=2}.\label{eqs11}
\end{equation}
which applies when $n_\mathrm{nose,2,g=2}p^{(+1)}(2\gamma_1)\ll1$. For $D=1$, when $n_\mathrm{nose,1,g=1}p^{(+1)}(\gamma_1)\ll1$ and $n_\mathrm{nose,1,g=2}p^{(+1)}(\gamma_1)\ll1$,
\begin{equation}
   Q_{\text{nose2gen},\gamma_1,D=1}
   \approx
    p^{(+1)}(\gamma_1)(sn_\mathrm{nose,1,g=1}+n_\mathrm{nose,1,g=2})
   .
\end{equation}
The first and second terms are the contributions from beneficial mutations at $g=1$ and $g=2$, respectively. This expression can be written in the same form as Eq.~\ref{eqs11},
\begin{equation}
    Q_{\text{nose2gen},\gamma_1,D=1}
   \approx 2\gamma_1
    p_+e^{-2\gamma_1}n'_\mathrm{nose,1}
\end{equation}
where
\begin{equation}
    n'_\mathrm{nose,1}=e^{\gamma_1}(sn_\mathrm{nose,1,g=1}+n_\mathrm{nose,1,g=2})/2.
\end{equation}
Comparing with Eq.~\ref{eqs11}, the two nose formation probabilities remain similar at the population level whenever the effective nose size $n'_\mathrm{nose,1}$ is similar to $n_\mathrm{nose,2,g=2}$. For a common starting nose size of one, $n'_{\mathrm{nose,1}}/n_\mathrm{nose,2,g=2}\approx 0.8$ at $\gamma_1 = 0.15$.

\section{VI. Corrections to theory of adaptation velocity}
In the main text, we computed the adaptation velocity from Eqs.~(1)--(4). This theory predicts slightly slower adaptation than observed in simulations because it neglects several higher-order effects. Here, we calculate the corresponding corrections, which bring the theory into close agreement with the simulations.

\subsection{A. Corrected new nose size}
In Eq.~(1) of the main text, we assumed that a newly generated nose initially consists of a single individual. We now account for the possibility that multiple individuals in the current nose acquire beneficial mutations simultaneously, so that the new nose can initially contain more than one individual.

Assuming independent mutational outcomes, the probability of generating a new nose of size $k$ is given by the binomial distribution
\begin{equation}
p_{\mathrm{nose},k}
=
\binom{n_{\mathrm{nose}}}{k}
(1-p^{(+1)})^{n_{\mathrm{nose}}-k}
(p^{(+1)})^k,
\quad k=1,\ldots,n_{\mathrm{nose}},
\end{equation}
where $n_{\mathrm{nose}}$ is the size of the current nose before mutation and $p^{(+1)}$ is the probability that an individual acquires a $+1$ mutation. The probability that a new nose is generated,
regardless of its initial size, is
\begin{equation}
P_{+1}
=
\sum_{k=1}^{n_{\mathrm{nose}}}p_{\mathrm{nose},k}
=
1-(1-p^{(+1)})^{n_{\mathrm{nose}}}.
\end{equation}
Conditioned on the generation of a new nose, the expected initial nose size is therefore
\begin{equation}
E(n_{\mathrm{nose,start}})
=
\frac{
\displaystyle\sum_{k=1}^{n_{\mathrm{nose}}}
k\,p_{\mathrm{nose},k}
}{
\displaystyle\sum_{k=1}^{n_{\mathrm{nose}}}
p_{\mathrm{nose},k}
}
=
\frac{n_{\mathrm{nose}}p^{(+1)}}
{1-(1-p^{(+1)})^{n_{\mathrm{nose}}}}.
\end{equation}

Including this correction, the expected nose size
after $g$ generations becomes
\begin{equation}
n_{\mathrm{nose}}(g)
=
E(n_{\mathrm{nose,start}})
2^g e^{-\gamma(1-p_+)(g/D-1)}.
\end{equation}

For example, for $\gamma=0.15$ and $p_+=1/300$,
we obtain $E(n_{\mathrm{nose,start}})\approx 1.24$
for $n_{\mathrm{nose}}=1024$, compared with
$E(n_{\mathrm{nose,start}})\approx 1.04$ for
$n_{\mathrm{nose}}=200$. Thus, this correction matters mainly when the current nose is large.

\subsection{B. Feeding by sub-nose}

Next, individuals in the sub-nose (i.e., with fitness one unit below the nose) can acquire a beneficial mutation and thereby join the nose. This feeding accelerates nose growth and hence adaptation.

To estimate the number of individuals that join the nose in this way, we need the size of the sub-nose over time. When a new nose forms, the previous nose becomes the sub-nose, so the initial sub-nose size $n_\mathrm{sub\text{-}nose,0}$ is the size of the previous nose at the moment it produced the new nose. We estimate the median of this size. While it was the nose, the previous nose grew by a factor $r=2^De^{-\gamma(1-p_+)}$ per selection cycle. Assuming it did not reach the cap $N$, the probability that it produced a new nose by the time it reached size $n_\mathrm{sub\text{-}nose,0}$ is
\begin{equation}
    P_\mathrm{nose,cf}
    =1-\prod_{j=0}^{\infty}
    (1-p^{(+1)})^{n_\mathrm{sub\text{-}nose,0}/r^j}
    =\frac{1}{2},
\end{equation}
where $p^{(+1)}$ is the probability that an individual has a $+1$ mutation.

Using $\ln(1-p^{(+1)})\approx-p^{(+1)}$ for $p^{(+1)}\ll1$, we obtain
\begin{equation}
    P_\mathrm{nose,cf}
    \approx 1-\exp\left[
    -n_\mathrm{sub\text{-}nose,0}p^{(+1)}
    \left(1+\frac{1}{r}+\frac{1}{r^2}+\cdots\right)
    \right]
    =\frac{1}{2}.
\end{equation}
The median sub-nose size when a new nose forms is thus
\begin{equation}
    n_\mathrm{sub\text{-}nose,0}
    =\frac{\ln 2\,(r-1)}
    {-r\ln(1-p^{(+1)})}
    \approx \frac{\ln 2\,(r-1)}{rp^{(+1)}}.
\end{equation}

The sub-nose subsequently grows approximately as $n_\mathrm{sub\text{-}nose}(g)=n_\mathrm{sub\text{-}nose,0}\,r^{g/D}$, and the number of individuals that move from the sub-nose into the nose in each selection cycle is $n_\mathrm{sub\text{-}nose}(g)\,p^{(+1)}$.

\subsection{C. Stochasticity of the nose}

Finally, the nose is lost if all of its individuals acquire deleterious mutations. Here, we estimate the importance of this effect.

For $D=1$, a new nose has two individuals after its first division. For $\gamma=0.2$, the probability that both acquire deleterious mutations is $(1-e^{-\gamma})^2=0.03$. Because the nose then grows by a factor of $2e^{-\gamma}$ per generation, the loss probability falls rapidly, to $4\times10^{-3}$ in the following generation. The nose is therefore rarely lost at this mutation rate. Thus, nose loss has only a minor effect on the adaptation velocity.

\section{VII. The case of mutating every division}
In the main text, we focused on the case in which mutations occur only after the last of the $D$ divisions, at rate $\gamma$. Here, we expand on the alternative case introduced in Fig.~3 of the main text, in which mutations occur after every division. So that $\gamma$ retains its meaning as the mean total number of mutations per selection cycle, mutations occur at rate $\gamma/D$ after each division. After the $D$ divisions, the top fraction $f=1/2^D$ of individuals is selected as before.

\begin{figure}[H]
    \centering

    % Panel (a)
    \begin{minipage}[t]{0.60\textwidth}
        \vspace{0pt}
        (a)\par\smallskip
        \includegraphics[width=\linewidth]{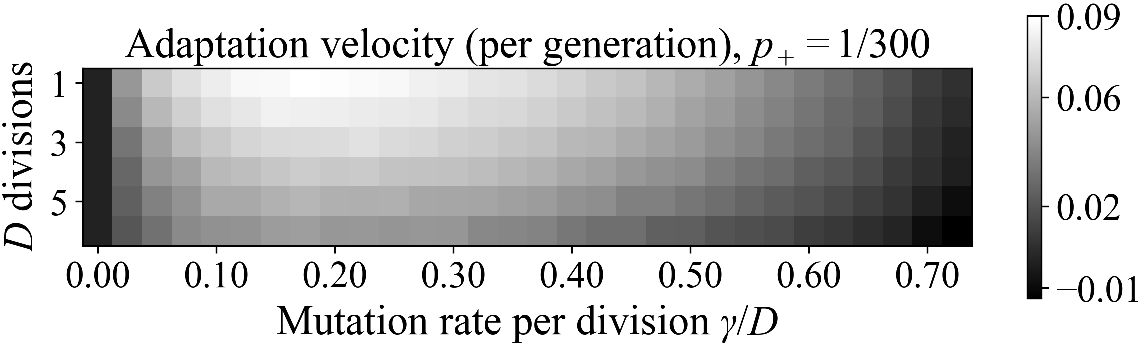}
    \end{minipage}
    \hfill
    % Panel (b)
    \begin{minipage}[t]{0.30\textwidth}
        \vspace{0pt}
        (b)\par\smallskip
        \includegraphics[width=\linewidth]{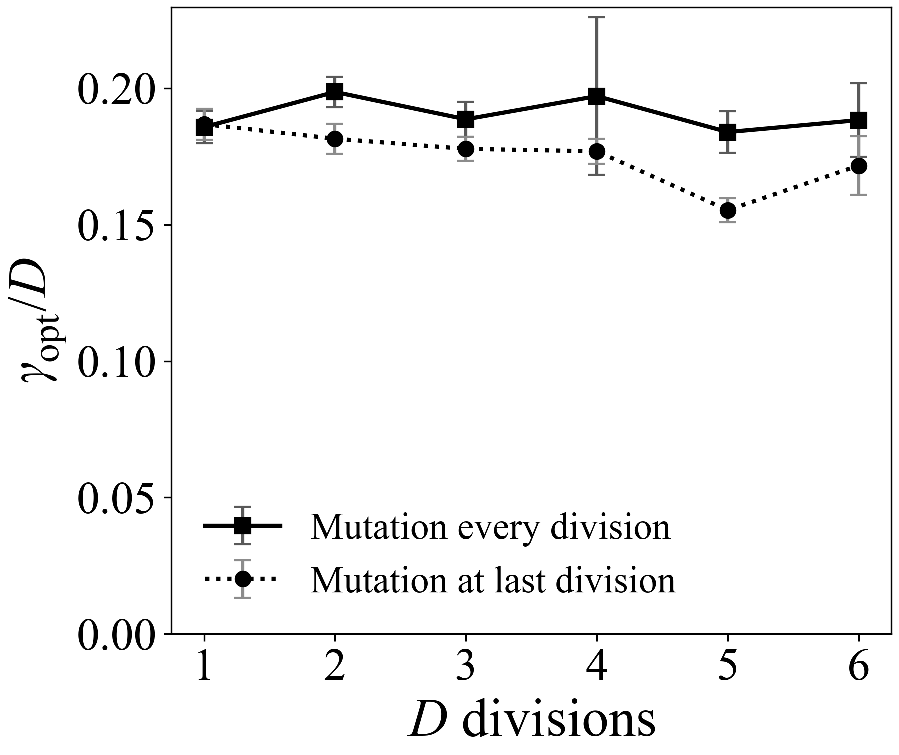}
    \end{minipage}

    \par\medskip

    % Panel (c)
    \begin{minipage}[t]{0.38\textwidth}
    \vspace{0pt}
    (c)\par
    \includegraphics[
        width=\linewidth,
        trim=0 0 0 1.5cm,
        clip
    ]{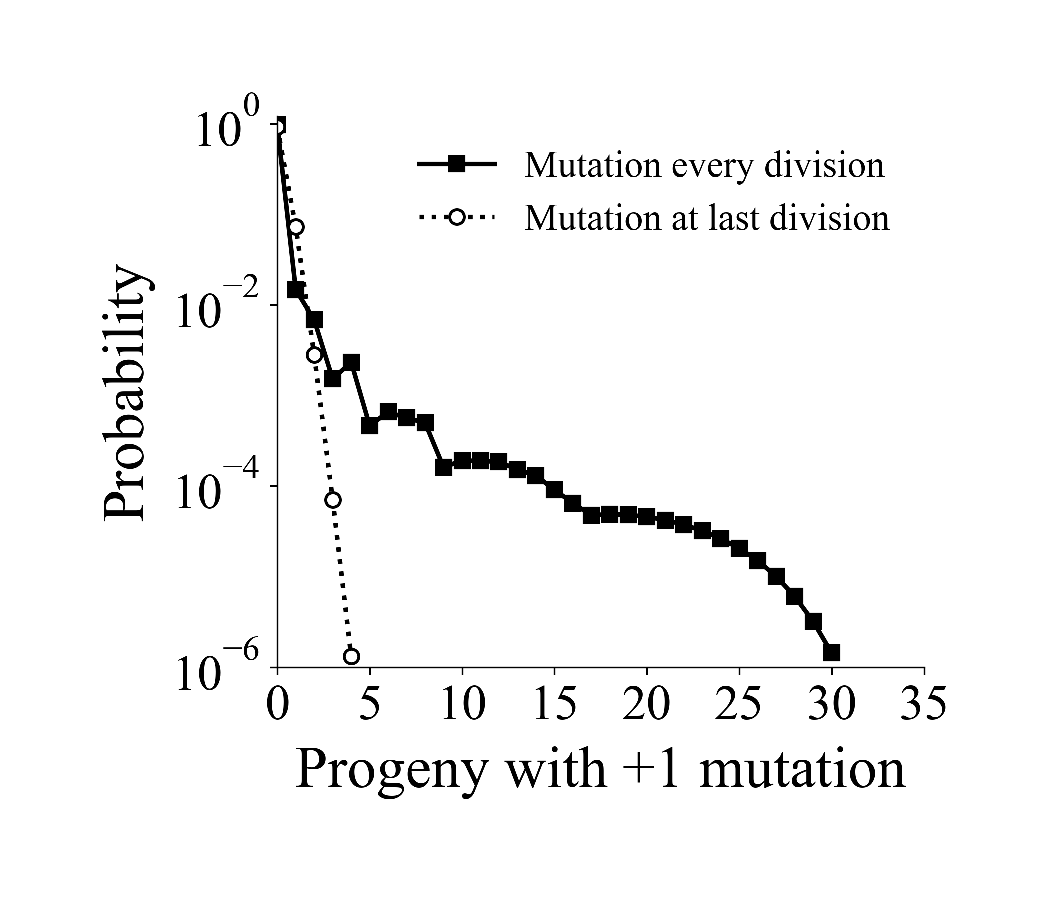}
\end{minipage}

    \caption{(a) Adaptation velocity for different numbers of divisions $D$ and mutation rates per division $\gamma /D$, for the case of mutating at every division. (b) Optimal mutation rates per division $\gamma _\mathrm{opt}/D$ across different $D$, comparing the cases of mutating at every division and mutating at the last division. (c) For $D=6$ and both mutational protocols, the distribution of the number of final progeny with $+1$ mutations (and no $-1$ mutations) obtained from a single progenitor over a cycle of $D$ divisions. Mutation rate per division is $\gamma /D=0.15$.}
    \label{fig:S1}
\end{figure}

We compare the two mutational protocols at the same $\gamma/D$. The optimal mutation rate per division, $\gamma_\mathrm{opt}/D$, is similar for the two protocols [Fig.~\ref{fig:S1}(b)]. However, the maximum adaptation velocity decreases more strongly with $D$ when mutations occur at every division [Fig.~3(a) of the main text].
In the main text, we attributed this difference to the lineage structure produced by the two protocols. When mutations occur at every division, a mutation arising early in the cycle is inherited by many descendants, producing a Luria--Delbr\"uck-like ``jackpot'' effect that increases clone sizes. Figure~3(b) of the main text shows the resulting clone-size distribution for the whole population, and Fig.~\ref{fig:S1}(c) shows it for a single progenitor. When mutations occur only at the last division, the number of beneficial progeny of a single progenitor is binomially distributed and well approximated by $\mathrm{Pois}(2^Dp_+\gamma e^{-\gamma })$, with mean $0.078$ for $D=6$ and $\gamma/D=0.15$ (i.e., $\gamma=0.9$). Mutating only at the last division is more likely to yield at least one beneficial progeny, even though the mean number of beneficial progeny is the same for both protocols.

We now show how this increased clonality slows adaptation. Because the adaptation velocity is set by the rate at which new noses form, we calculate the probability of new nose formation per selection cycle for both protocols. The calculation generalizes that of Sec.~V from two generations to $D$ divisions.

When mutations occur only at the last division, the probability that a new nose forms, i.e., that at least one individual in the nose acquires a beneficial mutation, is
\begin{equation}
    P_\mathrm{last}=1-(1-p^{(+1)}(\gamma ))^{n_\mathrm{nose}},
\end{equation}
where $n_\mathrm{nose}=2^Dn_0$ is the number of descendants, after $D$ divisions, of the $n_0$ nose individuals present at the start of the cycle. When mutations occur at every division, the mutation step that is followed by $k$ further divisions acts on $n_\mathrm{nose}/2^k$ individuals, and the probability that at least one beneficial mutation occurs during the cycle is
\begin{equation}
    P_{\mathrm{every,+1}}
    =
    1-\prod_{k=0}^{D-1}(1-p^{(+1)}(\gamma /D))^{n_\mathrm{nose}/2^k},
\end{equation}
As in Sec.~V, however, a beneficial mutation produces a new nose only if it is not cancelled by a deleterious mutation, both earlier along the same lineage and later in its descendants.

Let $s_k$ be the probability that an individual with $k$ divisions remaining has at least one descendant at the end of the cycle that has acquired no further mutations. Then
\begin{equation}
    s_0=1,
    \qquad
    s_{k+1}
    =
    1-\left(1-e^{-\gamma /D}s_k\right)^2.
\end{equation}
since each of the two daughters independently avoids mutation with probability $e^{-\gamma/D}$ and then has $k$ divisions remaining. The requirement that the lineage carried no deleterious mutation before the beneficial one contributes a further factor $e^{-(D-1-k)\gamma /D}$, the probability of no mutations in the $D-1-k$ earlier mutation steps. The probability of new nose formation is therefore
\begin{equation}
    P_{\mathrm{every}}
    \approx
    1-\prod_{k=0}^{D-1}
    \left[
        1-p^{(+1)}(\gamma /D)e^{-(D-1-k) \gamma /D}s_k
    \right]^{n_\mathrm{nose}/2^k}.
\end{equation}
For $n_\mathrm{nose}p^{(+1)}\ll 1$, these expressions simplify to
\begin{equation}
    P_\mathrm{last}\approx p^{(+1)}(\gamma )n_\mathrm{nose}
\end{equation}
for the case of mutating only at the last division, and to
\begin{equation}
    P_\mathrm{every}\approx p^{(+1)}(\gamma /D)e^{-(D-1)\gamma /D}n_\mathrm{nose}\sum _{k=0}^{D-1}e^{k\gamma /D}s_k/2^k \label{eq30}
\end{equation}
for mutation at every division. Using the identity $p^{(+1)}(\gamma)=Dp^{(+1)}(\gamma/D)\,e^{-(D-1)\gamma/D}$, the latter can be written as
\begin{equation}
    P_\mathrm{every}=p^{(+1)}(\gamma )n_\mathrm{nose}\eta =\eta P_\mathrm{last} \label{eqs31}
\end{equation}
where 
\begin{equation}
    \eta =\frac{1}{D}\sum _{k=0}^{D-1}e^{k\gamma /D}s_k/2^k.\label{eqs32real}
\end{equation}
is the factor by which mutating at every division reduces the probability of new nose formation. For $D=6$ and $\gamma /D=0.15$, $\eta =0.38$. Because $s_k\approx 1$, the reduction comes mainly from the factor $1/2^k$: a mutation arising $k$ divisions before the end of the cycle samples only $n_\mathrm{nose}/2^k$ individuals. This factor alone, $\eta\approx(1/D)\sum_{k}2^{-k}=0.33$, accounts for an approximately threefold reduction.

By contrast, deleterious mutations that cancel a beneficial one play only a minor role in the difference between the protocols: the factor $e^{-(D-1)\gamma /D}$ in Eq.~\ref{eq30} cancels when $P_\mathrm{every}$ is expressed in terms of $p^{(+1)}(\gamma)$. To make this explicit, $P_\mathrm{last}$ can be written in terms of $p^{(+1)}(\gamma /D)$ for direct comparison with Eq.~\ref{eq30},
\begin{equation}
    P_\mathrm{last}=1-(1-Dp^{(+1)}(\gamma /D)\cdot e^{-(D-1)\gamma /D})^{n_\mathrm{nose}}. \label{eqs32}
\end{equation}
This expression has an intuitive interpretation. By the additivity of Poisson random variables, a single mutation step at rate $\gamma$ is equivalent to $D$ consecutive steps at rate $\gamma/D$ applied to the same individual. Equation~\ref{eqs32} is thus the probability that at least one individual acquires exactly one beneficial mutation in one of these $D$ steps and no mutation in the others. Deleterious mutations therefore enter both protocols in the same way; the protocols differ only in how many individuals each mutation step samples. Since $p^{(+1)}(\gamma /D)\ll 1$,
\begin{equation}
    P_\mathrm{last}\approx Dp^{(+1)}(\gamma /D)\cdot e^{-(D-1)\gamma /D}{n_\mathrm{nose}}=P_\mathrm{every}/\eta
\end{equation}
consistent with Eq.~\ref{eqs31}.